\documentclass[12pt]{article}
\usepackage{graphicx}

\usepackage[usenames]{color}
\usepackage{makeidx}

\begin{document}

\title{Comment on {\it Local Unitary equivalence of multipartite pure qubits states}}
\author{Ana M. Martins \\
Instituto Superior T\'{e}cnico-IST, 
Universidade de Lisboa-UL, \\
1049-001
Lisboa, Portugal }



\maketitle





\begin{abstract} 
A Comment on the Letter by B. Kraus {\it Phys. Rev. Lett. }{\bf 104}, 020504 (2010).
\end{abstract}

\newpage

{\bf Comment on "Local Unitary equivalence of multipartite pure qubits states"}

Kraus \cite{Kraus2010} derives necessary and sufficient conditions for the equivalence of {\it arbitrary}, $n$-qbit pure states under Local Unitary (LU) operations. Theorem 1, stating that any two pure $n$-qubit states, with $\rho_i \neq \frac{\bf 1}{2}$, are LU-equivalent  iff their {\it standard forms} are equal, can be formulated using simply  the {\it trace decomposition} form of a state instead of its {\it standard form}, as proposed in the letter. By one side, it is simpler to work with the {\it trace decomposition form} of a state then with its {\it sorted trace decomposition}, that precedes its {\it standard form}. By another side we show that this new formulation is equally applicable to pure and mixed states. The proof goes like this.

A pure or mixed state  $\rho^{'}$ is LU-equivalent to $\rho$ when there exist Local Unitary operators $U_i , ..., U_n$ such that $\rho^{'}=U_1 \otimes ... \otimes U_n \rho \,\ U_n^{\dag} \otimes ... \otimes U_1^{\dag}$. Then 
\begin{equation}\label{1equiv}
 \rho_i^{'} = U_i \rho_i ,U_i^{\dag} 
 \end{equation}
where $ \rho_i^{'} $ and $ \rho_i$ are the reduced states of qubit $i$ associated to the global states $\rho^{'}$ and $\rho$. 

Let $V_i$ be the unitary operator that diagonalizes $\rho_i$, i.e., $\rho_i = V_i D_i V_i^{\dag}$ where $D_i$ is a diagonal matrix. Introducing this equality in eq.(\ref{1equiv}) we conclude that $V_i^{'} =U_i V_i$ is the unitary operator that diagonalizes $ \rho_i^{'}  $, i.e. $ \rho_i^{'} = V_i^{'} D_i V_i^{ ' \dag}$, therefore the  local unitary $U_i $ is uniquely determined (up to a global phase) by $U_i =V_i^{'} V_i^{\dag} $. When $ \rho_i^{'} =  \rho_i$, then $V_i^{'} =V_i$ and $U_i = {\bf 1}$, is the trivial identity operator. However, there are LU-operators, beside the trivial one $U_i = {\bf 1}$, named {\it cyclic} local unitary operators \cite{Fu2006,Gharibian2008}, that leave a state, $\rho_i \neq \frac {{\bf 1}}{2}$, invariant. Such LU operators obey the following commutation relation
\begin{equation}\label{cyclic}
[\rho_i , U_i] = 0   \,\,\   
 \end{equation}
The set of local unitary operators ${ U}_i \in SU(2)$  obeying the cyclic condition (\ref{cyclic}) is the stabilizer subgroup  of $\rho_i$. The general form of the non trivial unitary cyclic operators $U_j $,  that comute with the $1$-qubit density matrix $\rho_j = \frac{1}{2} ({\bf 1}_j + {\vec r} (j)  \cdot  {\vec \sigma (j)} )$, where ${\vec r} (j) \neq 0 $ is the 3-dimensional Bloch vector, $ {\vec \sigma} (j)  = ( \sigma_1 (j) , \sigma_2 (j), \sigma_3 (j)$, and where $\sigma_i (j)(i=1,2,3)$ are the Pauli matrices, was derived in Ref.\cite{Martins2013}. There it was shown that $U_j = e^{i \xi_j { \vec n_j} \cdot {\vec \sigma }(j)} $, where ${ \vec n_j} =  {\vec r} (j)/ || {\vec r} (j) ||$ and  $\xi_j$ is a real continuous parameter such that $0 \leq \xi_j \leq \pi/2 $. 

The {\it trace decomposition} $\rho_t$ and $\rho_t^{'}$ of {\it any} two LU-equivalent states $\rho$ and $\rho^{'}$ is
\begin{equation}\label{standard1}
\rho_t^{'} =  V_1^{' \dag} \otimes ... \otimes V_n^{' \dag} \,\   \rho^{'} \,\ V_n^{' } \otimes ... \otimes V_1^{' } 
 \end{equation}
 \begin{equation}\label{standard2}
\rho_t =  V_1^{\dag} \otimes ... \otimes V_n^{\dag} \,\   \rho \,\ V_n \otimes ... \otimes V_1
 \end{equation}
From where we easily obtain the relation between the two {\it trace decomposition} forms
\begin{equation}\label{relativon}
\rho_t^{'} =  (V_1^{' \dag}U_1 V_1) \otimes ... \otimes (V_n^{' \dag} U_n V_n ) \,\   \rho_t  \,\ (V_n^{' \dag} U_n V_n )^{\dag} \otimes ... \otimes (V_1^{' \dag}U_1 V_1)^{\dag}   
 \end{equation}
 When $ \rho_i^{'} \neq  \rho_i $ and $ \rho_i  \neq \frac {{\bf 1}}{2}$, for all $i=1,...,n$, then, $U_i =V_i^{'} V_i^{\dag} \Leftrightarrow V_i^{' \dag}U_i V_i = {\bf 1}_i $, and the two {\it trace decomposition} forms are equal, i.e., $\rho_t^{'} =\rho_t$. Given two arbitrary $n$-qubit states $\rho$ and $\rho^{'}$ we can check if they are LU-equivalent with the following protocol. After verifying that none of the LU operators is cyclic, we  express the density operators $\rho$ and $\rho^{'}$ in the generalized Pauli basis \cite{Martins2013}, and compute separately $\rho_t$ and $\rho^{'}_t$ using (\ref{standard1}) and (\ref{standard2}). Then we take the difference $(\rho^{'}_t -\rho_t )$. It will be null iff $\rho^{'} $ and $\rho$ are LU-equivalent. 
 
When there is at least one $k$ such that  $ \rho_k^{'} =  \rho_k   \neq \frac {{\bf 1}}{2}$, then ${\bar U}_k = V_k^{\dag}U_k V_k $ is given by
\begin{equation}\label{relation}
  {\bar U}_k =  \left(  \begin{array}{cc}
                             e^{i \omega_k} &  0 \\
                             0           &  e^{- i \omega_k} 
                             \end{array}
                           \right)
  \end{equation}
where $\omega_k = \xi_k || {\vec r} (k) ||$. Assuming that there is only one $U_k$ obeying condition (\ref{cyclic}) the {\it trace decomposition forms}, $\rho^{'}_t$ and $\rho_t$, are related by
\begin{equation}\label{relation2}
\rho_t^{'} =  {\bf 1}_1 \otimes ... \otimes {\bar U}_k \otimes ... \otimes {\bf 1}_n  \,\   \rho_t  \,\ {\bf 1}_1 \otimes ... \otimes {\bar U}_k^{\dag} \otimes ... \otimes {\bf 1}_n   
 \end{equation}
To check for the LU-equivalence we compute $\rho_t^{'}$ using eq.(\ref{standard1}) with $V_k^{'} = V_k$, and the right side of (\ref{relation2}) using (\ref{standard2}) and (\ref{relation}). Then we compute the difference, 
 \begin{equation}\label{relation3}
\rho_t^{'} - ( {\bf 1}_1 \otimes ... \otimes {\bar U}_k \otimes ... \otimes {\bf 1}_n  \,\   \rho_t  \,\ {\bf 1}_1 \otimes ... \otimes {\bar U}_k^{\dag} \otimes ... \otimes {\bf 1}_n )  
 \end{equation}
  It will be null iff $\rho^{'} $ and $\rho$ are LU-equivalent. This procedure is easily  generalized when there are more than one local unitary operador obeying condition (\ref{cyclic}). 

In conclusion, there are two advantages of using the {\it trace decomposition form} and density matrix representation of a state instead of the {\it  standard form} and the wave function representation of a state, as is done in the letter \cite{Kraus2010}: 1) It can be equally applied to pure and mixed states, 2) It is more straighforward to compute the {\it trace decomposition form} of a state than its {\it standard form} as it is proposed in the Letter.

\end{document}